\documentstyle[preprint,aps]{revtex}

\begin{document}

\draft

\preprint{    }

\title{Quantum Brans-Dicke Gravity in Euclidean Path Integral Formulation}

\author{Hongsu Kim}

\address{Department of Physics\\
Ewha Women's University, Seoul 120-750, Korea}

\date{November, 1996}

\maketitle

\begin{abstract}
 The conformal structure of Brans-Dicke gravity action is carefully
studied. It is discussed that Brans-Dicke gravity action has definitely no
conformal invariance. It is shown, however, that this lack of conformal 
invariance enables us to demonstrate that Brans-Dicke theory appears to
have a better short-distance behavior than Einstein gravity as far as 
Euclidean path integral formulation for quantum gravity is concerned.

\end{abstract}


\narrowtext
%\twocolumn

{\bf I. Introduction}
\\
In general, at the classical level (ignoring the quantum effects such as the
conformal anomaly that could break the conformal invariance
explicitly), one can always construct a conformally-invariant (Weyl-invariant)
matter field action. It is not clear, however, whether or not the gravity
sector of the action also needs to be conformally-invariant. For instance, if
one considers Einstein gravity, it
seems to suggest that the gravity action, in general, is not necessarily
conformally-invariant since the Einstein-Hilbert action is manifestly Weyl
non-invariant. 
Here in the present work, we are interested in the conformal nature of  
Brans-Dicke (BD) theory of gravity$^1$.
Since its conformal property has hardly been studied carefully$^2$,
we would like to take a closer look at its conformal structure in this work. 
We shall show that BD gravity action has no conformal invariance.
Remarkably, however, it will be shown that this lack of conformal invariance
enables us to demonstrate that the BD gravity theory possesses a much better
short-distance behavior than Einstein gravity.
The BD gravity action in the {\it Euclidean} signature takes the form
\begin{equation}
I_{E}[g, \Phi] = - \frac{1}{16 \pi} [\int_{M} d^{4}x \sqrt{g}
\{\Phi (R - 2 \Lambda)-
           \omega g^{\mu\nu} \frac{(\partial_{\mu} \Phi)(\partial_{\nu}
           \Phi)}{\Phi}\} + 2 \int_{\partial M} d^{3}x \sqrt{h}
           \Phi (K - K_{0})]
\end{equation}
where the ``BD $\omega$-parameter" is positive by definition and $K$ is the 
trace of
the second fundamental form of the boundary $\partial M$ and $h_{\mu\nu}$ 
is the
3-metric induced on $\partial M$. $K_{0}$ is
the trace of the second fundamental form of $\partial M$
embedded in flat spacetime. This constant term involving $K_{0}$ has
been added so that in cases when $\partial M$ is embedded in flat spacetime
the action of flat spacetime can be normalized to zero.
\\
{\bf II. Conformally Invariant ?}
\\
Upon examining the structure of BD gravity aciton, (but without the 
cosmological
constant, which is the potential term that explicitly breaks the classical
conformal invariance) one would naturally be tempted to the
possibility of conformal (Weyl) invariance and demand it
since the BD gravity action happens to have no dimensionful coupling constants
at all which is the ``bottom-line gualification" for any field theory action
to be scale invariant. In other words, in BD gravity action, the dimensionful
Newton's constant is replaced by the ``BD-scalar field", $G \rightarrow
1/\Phi(x)$ because that is the spirit of BD theory modification of
Einstein gravity, namely the realization of ``Mach's principle". And this
BD-scalar field $\Phi(x)$ is not allowed to enter the matter sector of the full
action to keep the success of ``principle of equivalence" 
which in turn leaves it
strictly massless regardless of energy scale. Besides, the kinetic term for the
BD-scalar field is divided by the BD-scalar itself making the undetermined
``BD $\omega$-parameter" dimensionless. In addition, the ``non-minimal" 
coupling of
BD-scalar field to the tensor, $\Phi R$ (here $R$ denotes the curvature
scalar) looks good enough to lead one to suspect
that it could play the central role in making the gravity action eventually 
Weyl
invariant similarly to the famous non-minimal coupling term, $\xi R \phi^{2}$
($\xi = \frac{1}{6}$ in 4-dim.) in conformally-invariant scalar matter field
theory. Therefore, BD gravity action appears to be in a good shape for being
Weyl-invariant.
Now, since the BD gravity action can be viewed as a kind of action for a scalar
(BD scalar) field theory coupled non-minimally to gravity (the metric tensor)
like this, to investigate the possibility of its being conformally-invariant 
let us first briefly review some of well-known features regarding the
``conformally-invariant field theories" (essentially matter field theories).
\hfill\break  Consider a field theory of a matter field $\psi_{\mu \nu ...}$
(which may be a spinor or tensor field) coupled Weyl invariantly to the
gravitational field, namely the theory of which the classical action is 
invariant under the Weyl rescaling of the metric and matter fields,
\begin{equation}
\{ x^{\mu}, ~g_{\mu \nu}(x), ~\psi_{\mu \nu ...}(x)\} \longrightarrow
\{ x^{\mu}, ~\Omega ^{2}(x)\tilde{g}_{\mu \nu}(x), ~\Omega ^{-d}(x)\tilde{\psi}
_{\mu \nu ...}(x)\}. 
\end{equation}
To be more concrete, for a scalar field the scale dimension (or conformal 
weight) 
is $d=(n-2)/ 2$, for a spinor (Dirac) field $d=(n-1)/ 2$ and for a vector
field in $n=4$ (and only in $4$) dimensions, $d=0$. 
Thus the scale dimensions of the scalar and spinor fields happen to coincide
with the mass (canonical) dimensions whereas it is not the case for the vector
field.
This is, of course, because  the underlying physics of Weyl rescaling (i.e. the
``local scale transformation") is not merely a dimensional analysis although   
they might seem to be identical.
In addition, for the Weyl
 invariance the action should be free of dimensionful coupling constants
(including the mass). Also, as is well-known, for scalar fields one needs to 
add the term ${1\over 2}\xi R\phi ^2$ (where $\xi=(n-2)/4(n-1)$ in
$n$-dimensions) in the action. And for  spinor fields one needs to introduce
a $n$-bein field which inherits its Weyl transformation law from the metric
field, i.e., $e^{a}_{\mu} \rightarrow \Omega (x)\tilde{e}^{a}_{\mu}$.
\hfill\break  Note that the essential requirement for the Weyl invariance is
the ``particular and appropriate" transformation law for the corresponding
matter field (i.e., ``field rescaling"), $\psi_{\mu\nu ...}(x) = 
\Omega ^{-d}(x) \tilde{\psi}_{\mu\nu ...}(x)$ (where $d$ is called 
``conformal weight") that accompanies the conformal
transformation of the metric, 
$g_{\mu\nu}(x) = \Omega ^{2}(x)\tilde{g}_{\mu\nu}(x)$.
\hfill\break  Here, however, it
should be emphasized that these field transformation laws are not strictly
implied (constrained) by the conformal transformation of metric itself, but
simply ``demanded" and then chosen that way for the conformal invariance.
Now going back to our main objective, in order to check the possibility of the
conformal invariance of BD gravity action, it appears necessary to view
the BD gravity action as the action of a scalar (BD scalar) field theory 
coupled
to gravity and then try to determine the field transformation law for the BD
scalar field that is required to accompany the conformal transformation of
metric for the conformal invariance, to start with.
 Since the scale dimension of the BD scalar
field would be $d= n/2$ because the naive mass dimension
of BD scalar field is $d_{0}[\Phi] = d_{0}[M^{2}_{pl}] = 2$ in $n = 4$-dim.,
one can readily
identify its field transformation law with $\Phi = \Omega^{-d}(x) 
\tilde{\Phi} = \Omega^{-2}(x)\tilde{\Phi}$ that accompanies the conformal 
transformation of the
metric, $g_{\mu\nu} = \Omega^{2}(x)\tilde{g}_{\mu\nu}$.  In fact, however,
we have been misled thus far.
Actually, attentive readers already must have noticed that
the BD gravity aciton can never be conformally-invariant. The pitfall is that
despite its seemingly qualified features such as the absence of dimensionful
coupling constants and the non-minimal coupling term, the relative sign between
the kinetic term for the BD scalar field and the non-minimal coupling term in
the BD gravity action has a ``wrong sign" compared to the usual relative sign
between the kinetic term and the conformal coupling term in the action for a
conformally-invariant scalar matter field. To illustrate this point, let us
write down the actions for above two cases in {\it Lorentzian} 
signature with sign convention, $g_{\mu\nu} = {\it diag}(-+++)$ and 
$R^{\alpha}_{\beta \mu \nu} = \partial_{\mu}\Gamma^{\alpha}_{\beta \nu}
-\partial_{\nu}\Gamma^{\alpha}_{\beta \mu} + 
\Gamma^{\alpha}_{\mu \lambda}\Gamma^{\lambda}_{\beta \nu }
- \Gamma^{\alpha}_{\nu \lambda}\Gamma^{\lambda}_{\beta \mu }$,   
\begin{eqnarray}
S_{BD} &=& \frac{1}{16 \pi} \int d^{4}x \sqrt{g} [\Phi R - \omega
           g^{\mu\nu}\frac{(\partial_{\mu} \Phi)(\partial_{\nu} \Phi)}{\Phi}],
           \nonumber \\
S_{M}  &=& - \int d^{4}x \sqrt{g} \frac{1}{2} [g^{\mu\nu}(\partial_{\mu}
           \phi)(\partial_{\nu} \phi) + \frac{1}{6}R \phi^{2}] \nonumber
\end{eqnarray}
where $S_{BD}$ is the BD gravity action and $S_{M}$ is the action for a
conformally-invariant scalar matter field. Pay special attention that Brans and
Dicke determined the sign of BD scalar field kinetic term in Lorentzian
signature such that with ``positive $\omega$", the contribution to the inertial
reaction (i.e. the variation of Newton's constant, $G = 1/\Phi(x)$) from the
nearby matter is positive$^1$, as it should be. In addition, there is, indeed,
another point of even more fundamental physical importance regarding the 
sign of
BD scalar field kinetic term from the field theory's point of view. That is, 
since the Euclidean action represents the energy of the system, the sign of
the BD scalar field kinetic term, or equivalently, the sign of $\omega$ in
the Euclidean BD gravity action given in eq.(1) should be positive.
If its sign were the other way around,
it would mean the negative kinetic energy of
BD scalar field which implies that the BD scalar field is a ``ghost field" with
negative norm and thus the BD gravity Lagrangian containing it violates
unitarity even at the tree level. The violation of unitarity of the BD gravity
theory is apparently even more damaging than its lack of conformal invariance.
(Einstein gravity does satisfy unitarity at the tree level$^3$.) 
Therefore, the relative sign between the BD scalar field kinetic term
and the non-minimal coupling term really has been determined correctly but it
happens to be ``opposite" to the relative sign between the kinetic term and the
conformal coupling term for the theory of a conformally invariant scalar matter
field. As a result, the BD gravity action simply cannot be invariant by its
nature under the conformal transformation. And actually,
one can explicitly show this conformal non-invariance of the BD gravity action
as follows.
 Obviously, the most straightforward way of checking the conformal invariance
of BD gravity action is
by carrying out the conformal transformation of BD gravity action and
seeing if it can be conformally-invariant under certain acceptable conditions.
Thus, it can be shown that under the conformal transformtions
\begin{equation}
g_{\mu\nu} = \Omega^{2}(x) \tilde{g}_{\mu\nu},~~   \Phi = \Omega^{-2}(x)
             \tilde{\Phi} \nonumber
\end{equation}
the Lorentzian BD gravity action transforms as
\begin{eqnarray}
S[g, \Phi] \longrightarrow &S&[\Omega, \tilde{g}, \tilde{\Phi}] 
          = \frac{1}{16 \pi} \int d^{4}x \sqrt{\tilde{g}}[\tilde{\Phi}
              \tilde{R} - \omega \tilde{g}^{\mu\nu} \frac{(\partial_{\mu}
              \tilde{\Phi})(\partial_{\nu} \tilde{\Phi})}{\tilde{\Phi}}
              \nonumber\\
          &-& 2(2 \omega + 3) \{\tilde{\Phi} \Omega^{-2} \tilde{g}^{\mu\nu}
              (\partial_{\mu} \Omega)(\partial_{\nu} \Omega) - \Omega^{-1}
              \tilde{g}^{\mu\nu} (\partial_{\mu} \Omega)(\partial_{\nu}
              \tilde{\Phi})\}].
\end{eqnarray}
Therefore again, one can notice that BD gravity action cannot be invariant 
under
the conformal transformations given in eq.(3) unless the BD 
$\omega$-parameter can
be chosen to be $\omega = - 3/2$ which is unacceptable. However, an additional
information one can extract out of this direct conformal transformation of BD
gravity action is that although it is not invariant under the local scale
transformations ($\Omega = \Omega (x)$), it can be invariant under the global
scale transformations ($\Omega =$ constant). Again, our general conclusion
is that BD gravity is not a conformally-invariant (or local Weyl invariant)
theory. 
Now, recall that the Einstein
gravity action itself is never conformally-invariant (not even global scale
invariant) either. Therefore, the conformal non-invariance of both 
Einstein and BD
gravity actions leads us to suspect that generally the gravity action does
not possess the conformal invariance.
\\
One might object our argument thus far by claiming that setting the BD
scalar field as $\Phi = \phi^2$, the BD gravity action given above now
takes the form
\begin{eqnarray}
S = {1\over 16\pi}\int d^4x \sqrt{g} [\phi^2 R - 4\omega g^{\mu\nu}
(\partial_{\mu} \phi)(\partial_{\nu} \phi)] \nonumber
\end{eqnarray}
which looks manifestly conformal invariant for suitable choice of $\omega$.
In order for this expression for the BD gravity action to be claimed to
be conformally invariant, it is required to choose $\omega = -3/2$ as is
obvious when compared with the conformally-coupled scalar theory case.
Again this is unacceptable since it demands a negative value of $\omega$
which, as emphasized, results in the negative-definite kinetic energy and
hence the violation of unitarity in the BD scalar sector of the action.
(This viewpoint of ours might not agree with that of reference$^2$. This,
however, is no contradiction since in ref. 2  they did not care much about 
the implication of the sign of $\omega$ and simply allowed $\omega$ to
have any sign.) 
\\
{\bf III. Exhibition of better short-distance behavior of BD gravity theory}
\\
According to the result of our preceding study, the BD
gravity theory has no conformal invariance by its nature. 
Thus under the conformal
transformation of metric, the field transformation law
for BD scalar field needs
not to be restricted by the conformal invariance condition and thus one has a
full freedom in choosing the transformation law for the BD scalar field. 
In order to convince ourselves of this point, it might be wise to remind
the definition of conformal transformation (or in a more appropriate term,
Weyl rescaling). A conformal transformation is the one that scales the 
metric by a spacetime position-dependent scalar factor. And the theory is
said to be conformally invariant if, when the metric is transformed in
this manner, there exists a set of transformations on the remaining fields
of the theory such that the action remains unchanged. And the term, Weyl
rescaling is usually used to indicate this simultaneous action of the
metric scaling and the field transformations. When a theory is not 
conformally invariant, however, there is no preferred choice for the field
transformation accompanying the metric scaling. Then one is free to choose
the transformation law for the non-metric fields.
Now
with these observations in mind in what follows we would like to show
explicitly that the BD theory appears to have a better short-distance
behavior than Einstein gravity as far as the Euclidean path integral
formulation for quantum gravity is concerned.
Again, consider the {\it Euclidean} BD-gravity
action including the cosmological constant term but dropping the irrelevant
surface term in eq.(1)
\begin{eqnarray}
I_{E}[g, \Phi] = - \frac{1}{16 \pi} \{\int_{M} d^{4}x \sqrt{g}[\Phi (R - 2
           \Lambda)- \omega g^{\mu\nu} 
         \frac{(\partial_{\mu} \Phi)(\partial_{\nu}\Phi)}{\Phi}]\}.
\end{eqnarray}
Now, if one reminds the essential point of the unboundedness of Einstein 
gravity
action, from the above expression for Euclidean BD gravity action one can 
readily
notice the possibility of its apparent boundedness from below. In other words,
first we note that the Euclidean gravity aciton represents gravitational 
energy.
Next, ignoring the surface term, in above expression for Euclidean BD gravity
action, the term proportional to the curvature scalar (non-minimal coupling
term) which essentially represents the kinetic energy of the tensor field can
become arbitrarily negative because the curvature, in general, can become
arbitrarily large and positive. Actually, this leads to the arbitrarily
negative gravitational energy in general relativity or equivalently
the unboundedness of Euclidean Einstein gravity action which makes the 
Euclidean
path integral for Einstein gravity essentially divergent. 
In the framework of BD gravity,
however, one more gravitational degree of freedom, namely the BD scalar field
$\Phi$ representing the variable Newton's constant has been added to the
theory. As a result, the kinetic term for the BD
scalar field in Euclidean BD gravity action contributes positive energy 
(as it
should just like the kinetic term of ordinary scalar matter field because
otherwise it would violate the unitarity as mentioned earlier) 
to the total gravitational energy
signaling the possibility that this positive contribution from the scalar part
might cancel or even overwhelm the arbitrarily negative contribution from the
tensor part of gravitational degrees of freedom. Therefore, in what follows, 
by employing the conformal transformation scheme discussed in the preceding
section, we shall show that the Euclidean path integral for BD gravity
exhibits milder divergence which may be interpreted as indicating better
short-distance behavior of the BD theory.
First,
let us examine the behavior of Euclidean BD gravity action under the conformal
transformation
\begin{eqnarray}
g_{\mu\nu} = \Omega^{2}(x) \tilde{g}_{\mu\nu}.
\end{eqnarray}
Since BD gravity action can never be conformally-invariant in the first place,
the choice of the accompanying field transformation law for the BD scalar field
does not necessarily have to be restricted by the conformal invariance
condition. In other words, one may now choose the accompanying transformation
law for the BD scalar field such that it transforms in an arbitrary (general)
conventional {\it linear} way
\begin{eqnarray}
\Phi = f(\Omega) \tilde{\Phi}.
\end{eqnarray}
Thus, under this Weyl rescaling, Euclidean BD gravity action,
in eq.(5) transforms as
\begin{eqnarray}
I_{E}[\Omega, \tilde{g}, \tilde{\Phi}]
  &=& - \frac{1}{16 \pi} \int d^{4}x \sqrt{\tilde{g}} f(\Omega) [\Omega^{2}
        \tilde{\Phi} \tilde{R} - \Omega^{4}\tilde{\Phi} 2 \Lambda - \Omega^{2}
        \omega \frac{(\partial_{\mu} \tilde{\Phi})(\partial^{\mu}
        \tilde{\Phi})}{\tilde{\Phi}} \nonumber \\
  & & + 6\{\tilde{\Phi} (1 + \Omega \frac{f'(\Omega)}{f(\Omega)})
        (\partial_{\mu} \Omega)(\partial^{\mu} \Omega) + \Omega(\partial_{\mu}
        \Omega)(\partial^{\mu} \tilde{\Phi})\}  \nonumber \\
  & & - \omega \{\tilde{\Phi}(\Omega \frac{f'(\Omega)}{f(\Omega)})^{2}
        (\partial_{\mu} \Omega)(\partial^{\mu} \Omega) + 2 \Omega (\Omega
        \frac{f'(\Omega)}{f(\Omega)}) (\partial_{\mu} \Omega)(\partial^{\mu}
        \tilde{\Phi})\}]
\end{eqnarray}
where $f'(\Omega)$ denotes the derivative of $f(\Omega)$ with respect to
$\Omega$.
Therefore, even at this stage, it appears that even if one chooses the 
``rapidly
varying" conformal factor $\Omega(x)$, there would be under a certain
circumstance no problem of ``negative
indefiniteness" of Euclidean gravity action or ``non-convergence" of Euclidean
path integral for gravity in the framework of BD theory of gravity because the
fluctuations of the conformal factor arising from the BD scalar field kinetic
term in the third line of eq.(8) may well overwhelm or at least cancel out 
those
arising from the kinetic term for tensor (i.e. the curvature scalar term,
$R$) in the second line. 
To be more explicit, let us cast the above conformal transformed Euclidean
BD gravity action to the following form
\begin{eqnarray}
I_{E}[\Omega, \tilde{g}, \tilde{\Phi}]
  &=& \frac{1}{16 \pi} \int d^{4}x \sqrt{\tilde{g}} f(\Omega) [- \Omega^{2}
      \tilde{\Phi} \tilde{R} + \Omega^{4}\tilde{\Phi} 2 \Lambda + \Omega^{2}
      \omega \frac{(\tilde{\nabla} \tilde{\Phi})^{2}}{\tilde{\Phi}} 
      \nonumber \\
  & & + \tilde{\Phi} \{\omega X^{2}(\Omega) - 6(1 + X(\Omega))\} 
      (\tilde{\nabla} \Omega)^{2} \nonumber \\
  & & + 2 \Omega \{\omega X(\Omega) - 3\} (\tilde{\nabla}\Omega)
  \cdot(\tilde{\nabla}\tilde{\Phi}) ]
\end{eqnarray}
where $X(\Omega) \equiv [\Omega \frac{f'(\Omega)}{f(\Omega)}]$. Now, since the
choice of conformal factor $\Omega(x)$ and the field transformation law for the
BD-scalar, $\Phi = f(\Omega) \tilde{\Phi}$ are arbitrary, if one chooses the
``rapidly varying" conformal factor $\Omega(x)$ and then the field 
transformation
law $f(\Omega)$ such that $\omega X^{2}(\Omega) - 6 (1 + X(\Omega)) > 0$
and that generally $(\tilde{\nabla} \Omega)^{2}$-term dominates over the
$(\tilde{\nabla}
\Omega)$-term, then the Euclidean BD gravity action can be made as arbitrarily
positive as desired.
Namely, for rapidly monotonously increasing, decreasing and rapidly
oscillating $\Omega(x)$, choose $f(\Omega)$ such that
\begin{equation}
\omega X^{2} - 6 (1 + X) > |\omega X - 3| > 0
\end{equation}
then it will do the job.
For instance, choose $f(\Omega) = \Omega^{-n}$ then
$\omega X^{2} - 6 (1 + X) = \omega n^{2} + 6 (n - 1)$ and
$|\omega X - 3| = \omega n + 3$, and the condition in eq.(10) can always be
satisfied by taking $n$ to be $n = {\rm positive ~integer} > 1$, regardless of
the value of $\omega$ (which, of course, is positive)! {\it Note that this 
choice
obviously includes the case when} $\Phi(x)$ {\it field follows the usual scale
transformation law}, $\Phi = \Omega^{-2} \tilde{\Phi}$.
Therefore, as we can see in the above example, one can always choose the 
conformal
factor $\Omega(x)$ and the field transformation law for the BD-scalar
$f(\Omega)$ such that the Euclidean BD gravity action can be made as large and
positive as wanted, even {\em regardless of the value of $\omega$-parameter for
a large class of the field transformation function $f(\Omega)$}.
Then
we may as a consequence conclude that the corresponding
Euclidean path integral formulation
for BD gravity can be well-defined. To see if this can be indeed the case,
we refer to the argument by Gibbons, Hawking and Perry$^4$ ; let $G$ be the 
space
of all compact, 4-metrics on a 4-dimensional manifold $M$. Now divide this
space $G$ into ``conformal equivalent classes of metrics" under the conformal
transformations, $g_{\mu\nu}(x) = \Omega^2(x)\tilde{g}_{\mu\nu}(x)$ 
(by conformal equivalent classes of metrics, it means that any
two metrics belonging to the same conformal equivalent class are related to
each other by a conformal transformation.). Then in performing the Euclidean
path integral for pure BD gravity  
\begin{eqnarray}
Z = \int [d\tilde{g}_{\mu\nu}][d\tilde{\Phi}][d\Omega] e^{-I_{E}
[\tilde{g}, \tilde{\Phi}, \Omega]}, 
\end{eqnarray}
in each conformal equivalent class, first
integrate over the conformal factors $\Omega(x)$ which is convergent as we
have seen in the above. Next, the remaining integral over different conformal
equivalent classes of metrics $\tilde{g}_{\mu\nu}$ would be convergent 
due to the ``positive action
theorem" by Schoen and Yau$^5$ and so would the integral over the conformal
transformed BD scalar field $\tilde{\Phi}(x)$.
\\
{\bf IV. Discussions}
\\
To conclude, by employing the conformal transformation scheme, we showed
that for a large class of the conformal transformations of BD scalar field,
the Euclidean BD gravity action can be shown to be bounded from below.
(Of course this analysis of ours is by no means a sound proof that the
Euclidean BD gravity action is indeed bounded below.)
This is in contrast to the case of Einstein gravity in which the same
conformal transformation scheme demonstrates that its Euclidean action
is unbounded below and hence its Euclidean path integral formulation is
ill-defined. Therefore this comparison between the two theories appears
to expose the fact that the BD gravity theory has a better short-distance
behavior since the typical quantities involved in quantum theory exhibit
milder divergences. And the BD scalar field which represents the variable
Newton's constant with space and time embodying the spirits of Mach's
principle and Dirac's large number hypothesis plays the central role in
making the theory better behaved at short-distance scales.
\\
\\
{\bf Acknowledgement}
\\
The author would like to thank prof. C. Goebel at the 
Univ. of Wisconsin-Madison for valuable discussions.
\\
\vspace{2cm}

{\bf  References}

\begin{description}
\item {1}  C.~Brans and R.~H.~Dicke, Phys. Rev. {\bf 124}, 925 (1961).
\item {2}  S.~Deser, Ann. Phys. (N.Y.) {\bf 59}, 248 (1970) ;
 P.~G.~O.~Freund, {\it ibid.} {\bf 84}, 440 (1974).
\item {3} 'tHooft and Veltman, Ann. Inst. Poincar\'{e} {\bf 20}, 69
(1974).
\item {4}  G.~W.~Gibbons, S.~W.~Hawking, and M.~J.~Perry, Nucl. Phys.
{\bf B138}, 141 (1978).
\item {5} R.~Schoen and S.T.~Yau, Phys. Rev. Lett. {\bf 42}, 547 (1979).

\end{description}
\end{document}